\documentclass[aps,prl,preprint,groupedaddress,longbibliography]{revtex4-1}
\usepackage{amssymb}
\usepackage{amsmath}
\usepackage{graphicx}
\usepackage{dcolumn}
\usepackage{lineno}

\begin{document}


\title{Towards Adoption of an Optical Second:\\ Verifying Optical Clocks at the SI Limit}


\author{W. F. McGrew$^{1,2}$, X. Zhang$^{1,*}$, H. Leopardi$^{1,2}$, R. J. Fasano$^{1,2}$, D. Nicolodi$^{1,2}$, K. Beloy$^{1}$, J. Yao$^{1,2}$, J. A. Sherman$^{1}$, S. A. Sch\"affer$^{1,\dagger}$, J. Savory$^{1}$, R. C. Brown$^{1,\ddagger}$, S. R\"omisch$^{1}$, C. W. Oates$^{1}$, T. E. Parker$^{1,\mathsection}$, T. M. Fortier$^{1,\mathparagraph}$, A. D. Ludlow$^{1,2,\|}$}
\affiliation{$^{1}$National Institute of Standards and Technology, 325 Broadway, Boulder, Colorado 80305, USA\\$^2$Department of Physics, University of Colorado, Boulder, CO 80309, USA}


\date{\today}

\begin{abstract}
The pursuit of ever more precise measures of time and frequency is likely to lead to the eventual redefinition of the second in terms of an optical atomic transition. To ensure continuity with the current definition, based on a microwave transition between hyperfine levels in ground-state $^{133}$Cs, it is necessary to measure the absolute frequency of candidate standards, which is done by comparing against a primary cesium reference. A key verification of this process can be achieved by performing a loop closure---comparing frequency ratios derived from absolute frequency measurements against ratios determined from direct optical comparisons. We measure the $^1$S$_0\!\rightarrow^3$P$_0$ transition of $^{171}$Yb by comparing the clock frequency to an international frequency standard with the aid of a maser ensemble serving as a flywheel oscillator. Our measurements consist of 79 separate runs spanning eight months, and we determine the absolute frequency to be 518 295 836 590 863.71(11) Hz, the uncertainty of which is equivalent to a fractional frequency of $2.1\times10^{-16}$. This absolute frequency measurement, the most accurate reported for any transition, allows us to close the Cs-Yb-Sr-Cs frequency measurement loop at an uncertainty of $<$3$\times10^{-16}$, limited by the current realization of the SI second. We use these measurements to tighten the constraints on variation of the electron-to-proton mass ratio, $\mu=m_e/m_p$. Incorporating our measurements with the entire record of Yb and Sr absolute frequency measurements, we infer a coupling coefficient to gravitational potential of $k_\mathrm{\mu}=(-1.9\pm 9.4)\times10^{-7}$ and a drift with respect to time of $\frac{\dot\mu}{\mu}=(5.3 \pm 6.5)\times10^{-17}/$yr.
\end{abstract}


\maketitle

\section{Introduction}

Since the first observation of the 9.2 GHz hyperfine transition of $^{133}$Cs, it was speculated that atomic clocks could outperform any conventional frequency reference, due to their much higher oscillation frequency and the fundamental indistinguishability of atoms \cite{Lyons1952}. Indeed, Harold Lyons' 1952 prediction that ``an accuracy of one part in ten billion may be achieved" has been surpassed one million-fold by atomic fountain clocks with systematic uncertainties of a few parts in $10^{16}$ \cite{Heavner2014}. The precision of atomic frequency measurements motivated the 1967 redefinition of the second in the International System of Units (SI), making time the first quantity to be based upon the principles of nature, rather than upon a physical artifact \cite{Terrien1968}. The superior performance of atomic clocks has found numerous applications, most notably enabling Global Navigation Satellite Systems (GNSS), where atomic clocks ensure precise time delay measurements that can be transformed into position measurements \cite{Ashby2002}.

Microwave atomic fountain clocks exhibit a quality factor on the order of $10^{10}$, and the current generation exhibits a line-splitting accuracy of a few parts in $10^6$. This leads to an uncertainty of several parts in $10^{16}$, i.e. the SI limit. Significant improvement of microwave standards is considered unrealistic; however, progress has been realized utilizing optical transitions, where the higher quality factor of approximately $10^{15}$ allows many orders of magnitude improvement \cite{Hall2006,Ludlow2015}. For example, a recent demonstration of two ytterbium optical lattice clocks at NIST found instability, systematic uncertainty, and reproducibility at the $1\times10^{-18}$ level or better, thus outperforming the current realization of the second by a factor of $>$100 \cite{McGrew2018}. The superior performance of optical clocks motivates current exploratory work aimed at incorporating optical frequency standards into existing time scales \cite{Ido2016,Grebing2016,Lodewyck2016,Hachisu2018,Yao2018,Yao2018a}. Furthermore, for the first time, the gravitational sensitivity of these clocks surpasses state-of-the-art geodetic techniques and promises to find application in the nascent field of chronometric leveling \cite{Delva2013}. Optical frequency references could potentially be standards not only of time, but of space-time.

Towards the goal of the eventual redefinition of the SI unit of time based on an optical atomic transition, the International Bureau of Weights and Measures (BIPM) in 2006 defined secondary representations of the second so that other transitions could contribute to the realization of the SI second, albeit with an uncertainty limited at or above that of cesium standards \cite{Gill2006}. Optical transitions designated as secondary representations (eight at the time of this writing) represent viable candidates for a future redefinition to an optical second, and the BIPM has established milestones that must be accomplished before adopting a redefinition \cite{Riehle2018}. Two key milestones are absolute frequency measurements limited by the $10^{-16}$ performance of cesium, in order to ensure continuity between the present and new definitions, and frequency ratio measurements between different optical standards, with uncertainty significantly better than $10^{-16}$. These two milestones together enable a key consistency check: it should be possible to compare a frequency ratio derived from absolute frequency measurements to an optically measured ratio with an inaccuracy limited by the systematic uncertainty of state-of-the-art Cs fountain clocks. Here we present a measurement of the $^{171}$Yb absolute frequency that allows a ``loop closure" consistent with zero at $2.4\times10^{-16}$, i.e., at an uncertainty that reaches the limit given by the current realization of the SI second.

\section{Experimental scheme}

This work makes use of the 578 nm $^1$S$_0\! \rightarrow^3$P$_0$ transition of neutral $^{171}$Yb atoms trapped in the Lamb-Dicke regime of an optical lattice at the operational magic wavelength \cite{Katori2003,Lemke2009}. The atomic system is identical to that described in Ref. \cite{McGrew2018} and has a systematic uncertainty of $1.4\times10^{-18}$. We note that only two atomic shifts (due to blackbody radiation and the second-order Zeeman effect) have a magnitude that is relevant for the $10^{-16}$ uncertainties of the present measurement. Several improvements have reduced the need to optimize experimental operation by reducing the need for human intervention. A digital acquisition system is used to monitor several experimental parameters. If any of these leave the nominal range, data is automatically flagged to be discarded in data processing. An algorithm for automatically reacquiring the frequency lock for the lattice laser was employed. With these improvements, an average uptime of 75\% per run was obtained during the course of 79 separate runs, distributed over eight months (November 2017 to June 2018).

\begin{figure}
\includegraphics[width=\textwidth,height=\textheight,keepaspectratio]{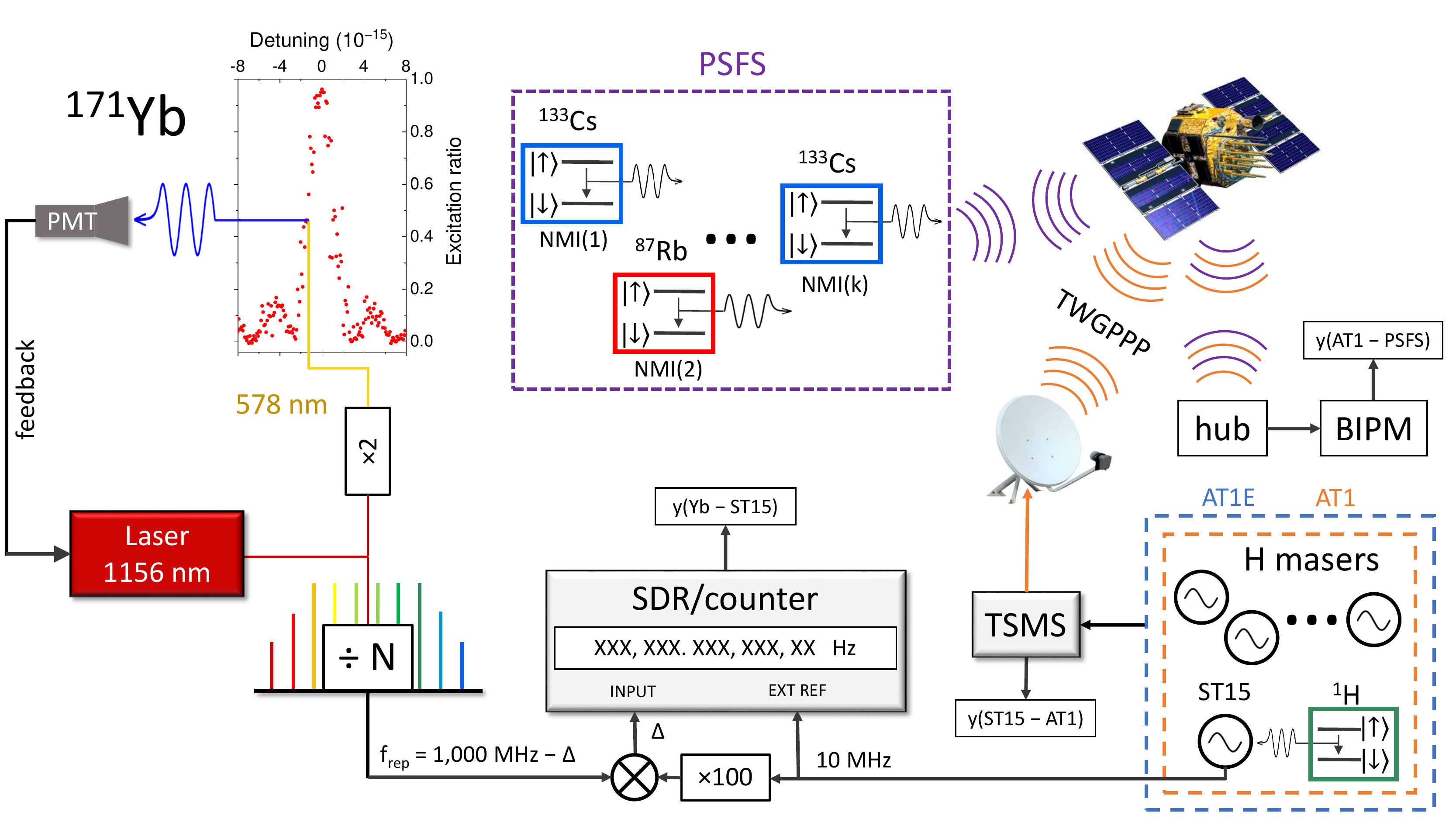}%
\caption{\label{setup} Experimental setup of the Yb optical lattice standard. A counter or software-defined radio (SDR) measures the beatnote between f$_\mathrm{rep}$ and the nominal 1 GHz reference derived from hydrogen maser ST15. The frequency of ST15 is compared by the NIST time scale measurement system (TSMS) to that of two maser ensembles---AT1E (blue) and AT1 (orange). These ensembles utilize the same masers (approximately eight, including ST15) but differ in the statistical weight given to each maser \cite{Parker1999}. The frequency of AT1 is sent to a central hub via the TWGPPP protocol \cite{Jiang2009}. The measurements are then sent from the hub to the BIPM by an internet connection, and the BIPM publishes data allowing a comparison of AT1 against Primary and Secondary Frequency Standard (PSFS), composed of $k$ separate clocks in different National Metrological Institutes (NMIs), where $k$ varies from five to eight during the measurements.}
\end{figure}

The experimental setup is displayed in Fig. \ref{setup}. A quantum-dot laser at 1156 nm is frequency-doubled and used to excite the 578 nm clock transition in a spin-polarized, sideband-cooled atomic ensemble trapped in an optical lattice. Laser light resonant with the dipole-allowed $^1$S$_0\! \rightarrow^1$P$_1$ transition at 399 nm is used to destructively detect atomic population, and this signal is integrated to apply corrections of the 1156 nm laser frequency so as to stay resonant with the ultranarrow clock line. Some of this atom-stabilized 1156 nm light is sent, via a phase-noise-cancelled optical fiber, to an octave-spanning, self-referenced Ti:sapphire frequency comb \cite{Fortier2006,Stalnaker2007}, where the optical frequency is divided down to $\mathrm{f}_\mathrm{rep}=1$ GHz$-\Delta$. This microwave frequency is mixed with a hydrogen maser (labelled here ST15), multiplied to a nominal 1 GHz, and the resultant $\Delta\approx300$  kHz heterodyne beatnote is counted.

The act of dividing the optical frequency down to 1 GHz may introduce systematic errors. Optical frequency synthesis introduces uncertainty that has been assessed to be well below $10^{-19}$, insignificant for the present experiment \cite{Yao2016,Leopardi2017}, but technical sources of error arising from the microwave setup may lead to inaccuracy. The nominal 10 MHz maser signal is multiplied by 100, to 1 GHz, by means of a frequency multiplier based on a phase-locked-loop. Electronic synthesis uncertainty is assessed by homodyne detection of the maser signal mixed with a 10 MHz signal generated by a direct digital synthesizer referenced to the 1 GHz signal. Electronic synthesis is found to contribute errors no larger than 3$\times10^{-17}$. Another source of uncertainty arises from counting error. The first half of the dataset is obtained using a ten-second-gated commercial frequency counter to count the heterodyne beatnote. Counting error is assessed by measuring the 10 MHz maser signal, also used as the counter's external reference. This counting error contributes an uncertainty of as much as $6\times10^{-14}$ of $\Delta$, leading to an error of $<2\times10^{-17}$ on f$_\mathrm{rep}$, and thus also on the optical frequency. The second half of the dataset is obtained replacing the counter with a software-defined radio (SDR) in two-channel differential mode \cite{Sherman2016}. The SDR phase-continuously measured the frequency once per second with zero dead time. The hardware acquisition rate and effective (software digital-filter) noise bandwidth were 1 MHz and 50 Hz, respectively. For all run durations the counting error of the SDR is $<1\times10^{-17}$ of f$_\mathrm{rep}$.

After the optical signal is down-converted and compared to the hydrogen maser ST15, the comb equation is used to determine a normalized frequency difference between the Yb optical standard and the maser, y(Yb$-$ST15). Throughout this work, we express normalized frequency differences between frequency standards A and B as follows,
\begin{widetext}
y(A$-$B)=$\frac{\nu_\mathrm{A}^\mathrm{act}}{\nu_\mathrm{A}^\mathrm{nom}}-\frac{\nu_\mathrm{B}^\mathrm{act}}{\nu_\mathrm{B}^\mathrm{nom}}\approx\frac{\nu_\mathrm{A}^\mathrm{act}/\nu_\mathrm{B}^\mathrm{act}}{\nu_\mathrm{A}^\mathrm{nom}/\nu_\mathrm{B}^\mathrm{nom}}-1$,
\end{widetext}
where $\nu^\mathrm{act(nom)}_\mathrm{X}$ is the actual (nominal) frequency of standard X, and the approximation is valid in the limit $(\nu^\mathrm{act}_\mathrm{X}-\nu^\mathrm{nom}_\mathrm{X})/\nu^\mathrm{nom}_\mathrm{X}\ll1$, a well-founded assumption throughout this work. In the definition of y(Yb$-$ST15), $ \nu_\mathrm{Yb}^\mathrm{nom}=\nu_\mathrm{Yb}^\mathrm{BIPM17}$ = 518 295 836 590 863.6 Hz is the 2017 BIPM recommended frequency of the Yb clock transition \cite{Riehle2018} and $\nu_\mathrm{ST15}^\mathrm{nom}=10$ MHz. The NIST time scale measurement system is used to transfer the frequency difference, y(Yb$-$ST15), from maser ST15 to a local maser ensemble significantly more stable than ST15. The ensemble serves as a flywheel oscillator for a comparison to an average of Primary and Secondary Frequency Standards (PSFS), which the BIPM publishes with a resolution of one month in Circular T \cite{circT}. The dead time uncertainty associated with intermittent operation of the optical standard is comprehensively evaluated in Appendix A and amounts to the largest source of statistical uncertainty; see Table \ref{tab:table1}. The maser ensemble frequency is transmitted to the BIPM via the hybrid Two-Way Satellite Time and Frequency Transfer/GPS Precise Point Positioning (TWGPPP) frequency transfer protocol \cite{Jiang2009}, and the frequency transfer uncertainty is the second-largest source of statistical uncertainty. The transfer process from the local maser ensemble to PSFS is described in Appendix B. The frequency transfer processes from ST15 to the local maser ensemble and finally to PSFS are continually operating, thus transferring the frequencies between the standards with no dead time.

 \begin{table}
 \caption{\label{tab:table1} Uncertainty budget of the eight-month campaign for the absolute frequency measurement of the $^{171}$Yb clock transition. A detailed discussion of the type A uncertainties is found in Appendices A and B. Data for March 2018 are shown as an example of one month's data.}
 \begin{ruledtabular}
 \begin{tabular}{lccc}
Uncertainty ($10^{-16}$) & March 2018 & Full campaign \\
\colrule
\textbf{Type A uncertainty}\\
\hspace{10 mm}Dead time & 2.5 & 1.0\\
\hspace{10 mm}Frequency transfer & 2.6 & 0.9\\
\hspace{10 mm}Yb-maser comparison & 0.8 & 0.4\\
\hspace{10 mm}Time scale measurement & $<$0.1 & $<$0.1\\
\hspace{10 mm}PSFS & 1.4 & 0.5\\
\textbf{Total type A} & \textbf{4.0} & \textbf{1.6}\\
\colrule
\textbf{Frequency comb type B uncertainty}\\
\hspace{10 mm}Optical synthesis & $<$0.001 & $<$0.001\\
\hspace{10 mm}Electronic synthesis & 0.3 & 0.3\\
\hspace{10 mm}Counter/SDR & 0.1 & 0.1\\
\textbf{Total comb type B} & \textbf{0.3} & \textbf{0.3}\\
\colrule
\textbf{PSFS type B} & \textbf{1.4} & \textbf{1.3}\\
\colrule
\textbf{Yb type B} & \textbf{0.014} & \textbf{0.014}\\
\colrule
\textbf{Relativistic redshift} & \textbf{0.06} & \textbf{0.06}\\
\hline \hline
\textbf{\textit{Total}} & \textbf{{4.2}} & \textbf{{2.1}}\\
 \end{tabular}
 \end{ruledtabular}
 \end{table}

\section{Results and analysis}

\begin{figure}
\includegraphics[width=\textwidth,height=\textheight,keepaspectratio]{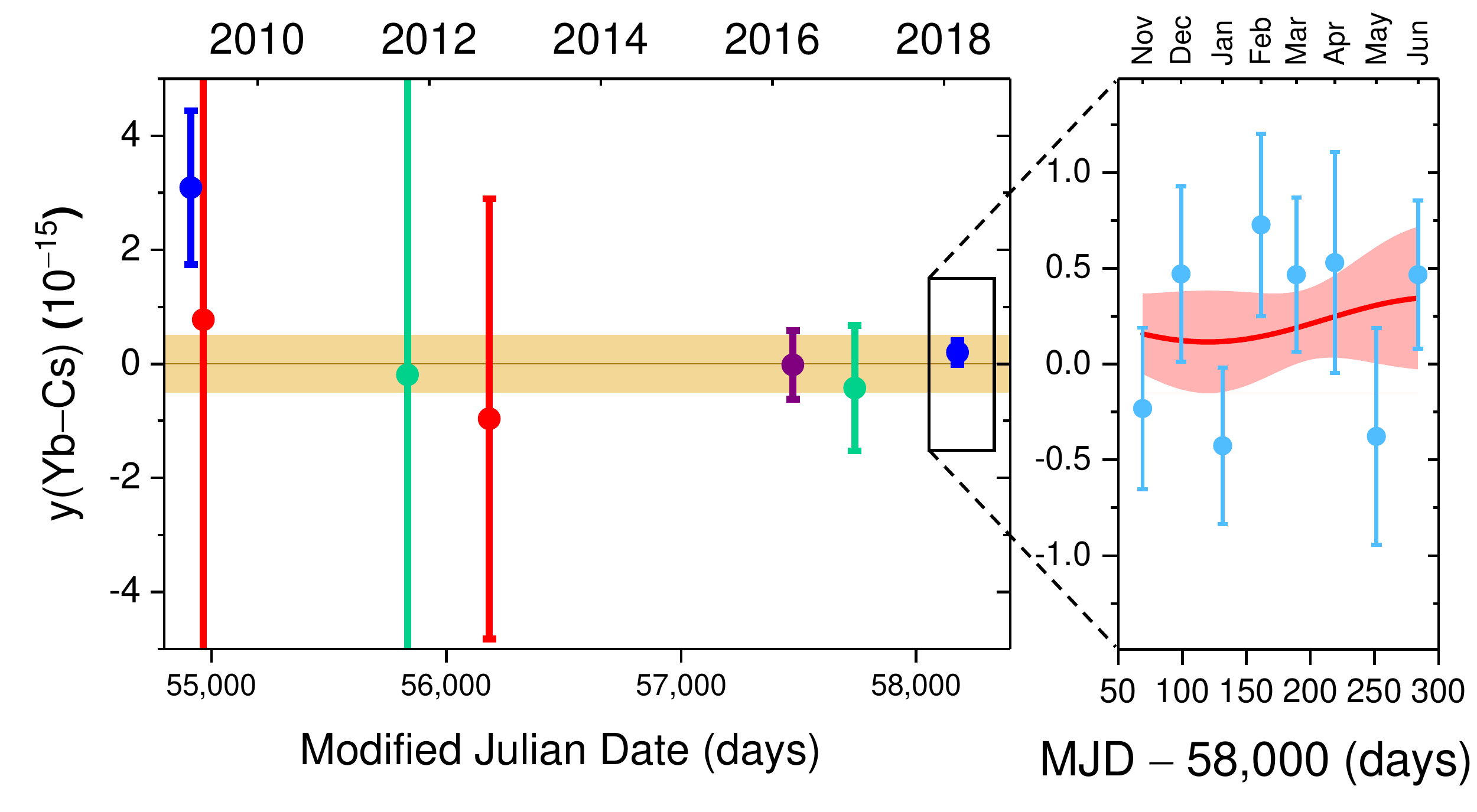}
\caption{\label{ybhistrec} Absolute frequency measurements of the $^1$S$_0\! \rightarrow^3$P$_0$ transition frequency measured by four different laboratories: NIST (blue) \cite{Lemke2009}, NMIJ (red) \cite{Kohno2009, Yasuda2012}, KRISS (green) \cite{Park2013, Kim2017}, and INRIM (purple) \cite{Pizzocaro2017}. The light-blue points in the inset represent the eight monthly values reported in this work, y$_\mathrm{m}$(Yb$-$PSFS), and the final dark-blue point represents y$_\mathrm{T}$(Yb$-$PSFS). The yellow shaded region represents the 2017 BIPM recommended frequency and uncertainty.  The inset shows a sinusoidal fit of the coupling parameter to gravitational potential for measurements of the frequency ratio between Yb and Cs between November 2017 and June 2018. The red shaded region in the inset represents 1$\sigma$ uncertainty in the fit function. }
\end{figure}

We make 79 measurements over the course of eight months, for a total measurement interval of 12.1 days, or a 4.9\% effective duty cycle. The weighted mean of the eight monthly values, y$_\mathrm{m}$(Yb$-$PSFS), gives a value for the total normalized frequency difference obtained from these measurements, y$_\mathrm{T}$(Yb$-$PSFS) and its associated uncertainty. The statistical (type A) and systematic (type B) uncertainties are accounted for in Table \ref{tab:table1}. Type B uncertainties tend to be highly correlated over time and therefore do not average down with further measurement time. Following convention, here we treat the type B uncertainties of the PSFS ensemble's constituent standards as uncorrelated between standards, enabling a PSFS type B uncertainty of $1.3\times10^{-16}$, lower than the uncertainty of any individual fountain. We measure a value of $\nu_\mathrm{Yb} =$ 518 295 836 590 863.71(11) Hz, which represents a fractional frequency difference of $(2.1\pm2.1)\times10^{-16}$ from the 2017 BIPM recommended value of the Yb frequency \cite{Riehle2018}. The reduced-chi-squared statistic, $\chi^2_\mathrm{red}$, is 0.98, indicating that the scatter in the eight monthly values is consistent with the stated uncertainties. This represents the most accurate absolute frequency measurement yet performed on any transition. Furthermore, good agreement is found between this measurement and previous absolute frequency measurements of the Yb transition (Fig. \ref{ybhistrec}). If a line is fit to our data, the slope is found to be $(2.0\pm2.2)\times10^{-18}$/day, indicating that there is no statistically significant frequency drift.

Due to the unavailability of a local Cs primary frequency reference during this period, these measurements were performed without one. This mode of operation limits the achieveable instability---with a local Cs fountain clock and a low-instability microwave oscillator, it is possible to achieve type A uncertainties at the low $10^{-16}$ level after one day of averaging, whereas in our configuration this was not achieved until $>$10 days of cumulative runtime. Furthermore, it is necessary to correctly account for dead time uncertainty, as frequency measurements of the maser ensemble against PSFS are published on a very coarse grid. On the other hand, the unprecedented accuracy reported in this work is directly facilitated by the lower type B uncertainty associated with the PSFS ensemble, as compared with any single Cs fountain. An additional advantage to this mode of operation is that it is straightforward to determine frequency ratios with other secondary representations of the second that may be contributing to PSFS. For example, during these measurements, a Rb fountain clock (SYRTE FORb) contributed to PSFS \cite{Guena2014}, allowing the first direct measurement of the Yb/Rb ratio, found to be $\nu_\mathrm{Yb}/\nu_\mathrm{Rb}=$ 75 833.197 545 114 192(33); see Appendix C.

 \begin{figure}

\includegraphics[width=\textwidth,height=\textheight,keepaspectratio]{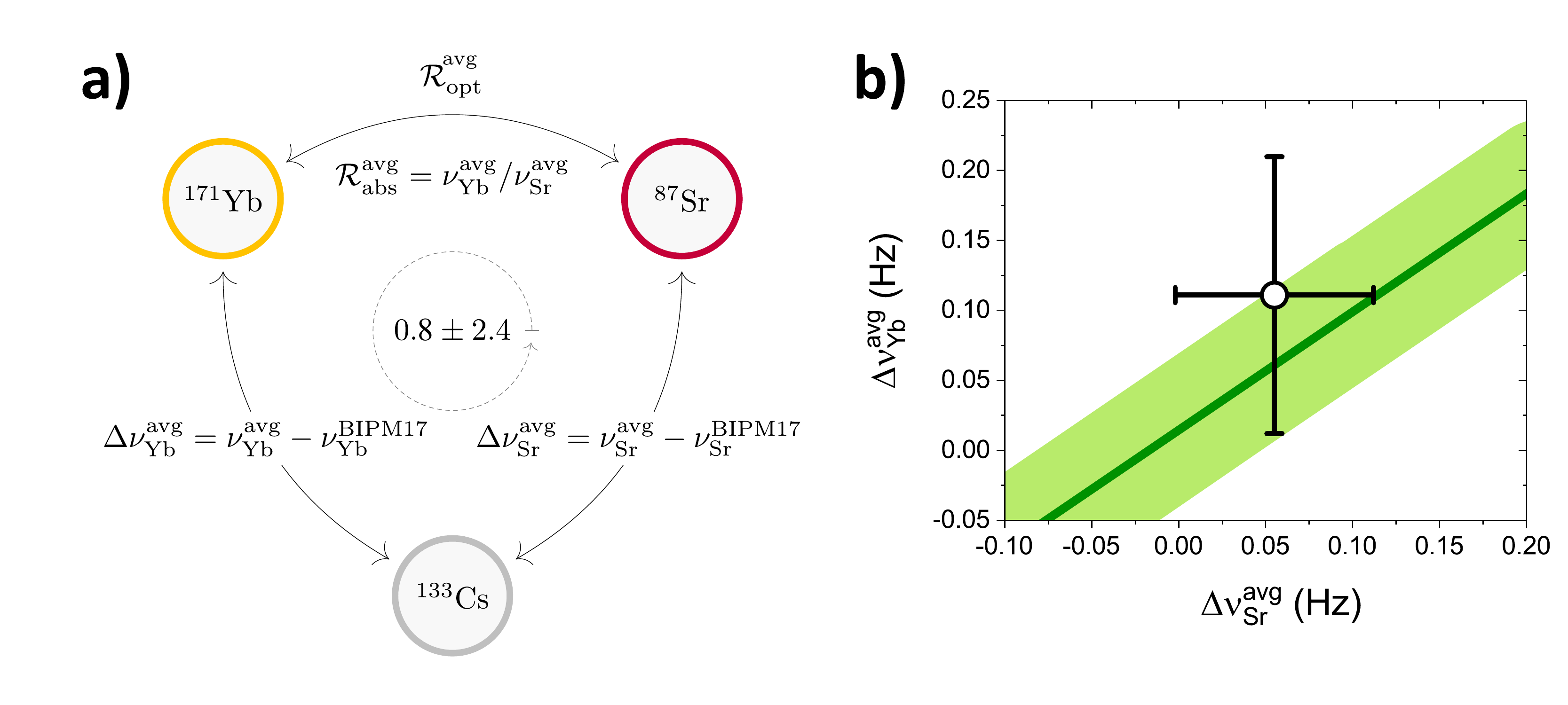}%
\caption{\label{loopclosure} Graphical representation of the agreement between frequency ratios derived from absolute frequency measurements of $^{171}$Yb and $^{87}$Sr and direct optical measurements. a) Schematic of the Cs-Yb-Sr-Cs loop that is examined. The central number is the misclosure, as parts in $10^{16}$. b) Average Yb and Sr frequency, parametrically plotted against each other. The error bars are the 1$\sigma$ uncertainty in the averaged absolute frequency measurements. The optical ratio measurement (dark green) appears as a line in this parameter-space, with the shaded region representing the uncertainty of the ratio. Frequency ratios derived from absolute frequencies agree well with ratios measured optically.}
\end{figure}

It is desirable to establish the consistency of frequency ratios determined through direct comparisons and through absolute frequency measurements. For absolute frequencies, the BIPM recommended values are based upon a least-squares algorithm that takes as inputs both absolute frequency measurements, as well as optical ratio measurements \cite{Margolis2015,Riehle2018}. To establish the consistency between absolute frequency measurements and direct optical ratio measurements, we determine average frequencies from the former only, as a weighted average of all previous measurements. If $\chi^2_\mathrm{red}>1$, we expand the uncertainty of the mean by $\sqrt{\chi^2_\mathrm{red}}$. For the Yb frequency, we determine a weighted average of the present work and six previous measurements \cite{Lemke2009,Kohno2009,Yasuda2012,Park2013,Kim2017,Pizzocaro2017}, $\nu_\mathrm{Yb}^\mathrm{avg}=$ 518 295 836 590 863.714(98) Hz. For the Sr frequency, we likewise determine a weighted average of 17 previous measurements  \cite{Ludlow2006,LeTargat2006,Boyd2007,Baillard2008,Gretchen2008,Hong2009,Falke2011,Yamaguchi2012,LeTargat2013,Akamatsu2014a,Falke2014,Tanabe2015,Lin2015,Lodewyck2016,Grebing2016,Hachisu2017,Hachisu2017a}, $\nu_\mathrm{Sr}^\mathrm{avg}=$ 429 228 004 229 873.055(58) Hz. The frequency ratio derived from absolute frequency measurements is therefore, $\mathcal{R}_\mathrm{abs}^\mathrm{avg}=\nu_\mathrm{Yb}^\mathrm{avg}/\nu_\mathrm{Sr}^\mathrm{avg}=$ 1.207 507 039 343 337 86(28). A frequency ratio can also be determined directly from optical frequency ratio measurements. From a weighted average of six optical ratio measurements \cite{Akamatsu2014,Takamoto2015,Nemitz2016,Grotti2018,Akamatsu2018,Fujieda2018}, we determine $\mathcal{R}_\mathrm{opt}^\mathrm{avg}=$ 1.207 507 039 343 337 768(60). We therefore determine a loop misclosure of ($\mathcal{R}_\mathrm{abs}-\mathcal{R}_\mathrm{opt}) / \mathcal{R}=(0.8\pm2.4)\times10^{-16}$, indicating consistency between the optical and microwave scales at a level that is limited only by the uncertainties of Cs clocks. This agreement is demonstrated graphically in Fig. \ref{loopclosure}. We emphasize that each of the three legs of the loop---Yb absolute frequency, Sr absolute frequency, and Yb/Sr ratio---feature different measurements performed at multiple laboratories across the world and are thus largely uncorrelated from each other. 

\section{New limits on coupling of $m_e/m_p$ to gravitational potential}

Many beyond-Standard-Model theories require that parameters traditionally considered fundamental constants may vary across time and space \cite{Safronova2017}. This hypothesized variation is detectable by looking for a change in the frequency ratio of two different types of atomic clock \cite{Flambaum2006}. We analyze our eight-month frequency comparison data to place bounds upon a possible coupling of the measured Yb/Cs frequency ratio to the gravitational potential of the Sun. We fit our data to y(Yb$-$PSFS) $= A$ $\mathrm{ cos}(2\pi(t-t_0)/1$ yr$)+y_0$, where $A$ and $y_0$ are free parameters, $t$ is the median date for each of the eight months, $t_0$ is the date of the 2018 perihelion, and 1 yr $=365.24$ days is the mean length of the tropical year. From our data, we determine the yearly variation of the Yb/Cs ratio, $A_\mathrm{Yb, Cs}=(-1.3\pm2.3)\times10^{-16}$; see the inset to Fig. \ref{ybhistrec}. The amplitude of the annual variation of the gravitational potential is $\Delta\Phi=(\Phi_\mathrm{max}-\Phi_\mathrm{min})/2\approx(1.65\times10^{-10}) \mathrm{c}^2$, where c is the speed of light in vacuum. Therefore, the coupling of the Yb/Cs ratio to gravitational potential is given by $\beta_\mathrm{Yb,Cs} = A_\mathrm{Yb,Cs}/(\Delta\Phi/\mathrm{c}^2)=(-0.8\pm1.4)\times10^{-6}$. A non-zero $\beta$ coefficient would indicate a violation of the Einstein equivalence principle, which requires that the outcome of any local experiment (e.g., a frequency ratio measurement) is independent of the location at which the experiment was performed. No violation of the equivalence principle is observed.

 \begin{table}
 \caption{\label{tab:table2} Measurements of coupling of dimensionless constants to gravitational potential. Sensitivity coefficients, $\Delta$K, are from \cite{Flambaum2006,Flambaum2009,Dinh2009}.}
 \begin{ruledtabular}
 \begin{tabular}{llccccc}
No. & Reference & X, Y & $\beta_\mathrm{X, Y}$ $(10^{-6})$ & $\Delta K^\alpha_{X,Y}$ & $\Delta K^{X_q}_{X,Y}$ & $\Delta K^{\mu}_{X,Y}$\\
\colrule
(i) & Dzuba \& Flambaum, 2017 \cite{Dzuba2017} & Al$^+$, Hg$^+$ & $0.16\pm0.30$ & 2.95 & 0 & 0 \\
(ii) & Ashby et al., 2018 \cite{Ashby2018} &  H, Cs & $ 0.22 \pm 0.25 $ & -0.83 & -0.102 & 0 \\
(iii) & This work & Yb, Cs & $-0.8 \pm 1.4$ &  -2.52 & -0.002 & -1 \\
 \end{tabular}
 \end{ruledtabular}
 \end{table}

Were this violation to occur, it might arise due to variation of the fine structure constant, $\alpha$; the ratio of the light quark mass to the quantum chromodynamics (QCD) scale, $X_q=m_q/\Lambda_{QCD}$; or the electron-to-proton mass ratio, $\mu=m_e/m_p$. To discriminate between each of these constants, we combine our results with two previous measurements---an analysis \cite{Dzuba2017} of a prior optical-optical measurement \cite{Rosenband2008} and a microwave-microwave measurement \cite{Ashby2018}. These results are chosen as they exhibit sensitivities to fundamental constants that are nearly orthogonal to each other and to our optical-microwave measurement. Table \ref{tab:table2} displays the coupling to gravitational potential observed in each measurement, as well as the differential sensitivity parameter $\Delta K_{X,Y}^\epsilon$, defined by $\delta$y(X$-$Y)$=\sum\limits_{\epsilon}\Delta K_{X,Y}^\epsilon(\delta\epsilon/\epsilon)$, where X and Y are the two atomic clocks being compared, and $\epsilon$ is $\alpha$, $X_q$, or $\mu$. We first use line (i) of Table \ref{tab:table2} to constrain the coupling parameter of $\alpha$, $k_\alpha=\beta_\mathrm{Al^+,Hg^+}/\Delta K^\alpha_\mathrm{Al^+,Hg^+}=(0.5\pm1.0)\times10^{-7}$. Applying this coefficient to line (ii) and propagating the errors, we find $k_{X_q}=(-2.6\pm2.6)\times10^{-6}$. Applying both of these coefficients to the present work in line (iii), we obtain a coupling coefficient to gravitational potential of $k_{\mu}=(0.7\pm1.4)\times10^{-6}$. This value represents an almost fourfold improvement over the previous constraint, $k_{\mu}=(-2.5\pm5.4)\times10^{-6}$ \cite{Peil2013}. In Appendix D, we extend our analysis to the full record of all Yb and Sr absolute frequency measurements to infer $k_{\mu}=(-1.9 \pm 9.4)\times10^{-7}$ and $\frac{\dot\mu}{\mu}=(5.3 \pm 6.5)\times10^{-17}$/yr.

\section{Conclusions}

We have presented the most accurate spectroscopic measurement of any optical atomic transition. We find that the frequency ratio derived from $^{171}$Yb and $^{87}$Sr absolute frequency measurements agrees with the optically measured ratio at a level that is primarily limited by the uncertainties of state-of-the-art fountain clocks. This level of agreement bolsters the case for redefinition in terms of an optical second. Further progress can be realized by the closing of loops consisting exclusively of optical clocks, since the improved precision of these measurements will allow misclosures that are orders of magnitude below the SI limit.

\appendix
\section{Appendix A: Frequency connection to AT1E and dead time uncertainty}

After determination of the normalized fractional frequency difference y(Yb$-$ST15), the difference between ST15 and the AT1E and AT1 maser ensembles is measured, where AT1E is a post-processed time scale that uses the same algorithm and clock measurement data as the real-time time scale AT1 (identified by the BIPM as TA(NIST)) \cite{Parker1999}. Since AT1E is post-processed, it has generally better stability than AT1 because clocks suffering from environmental perturbations and other technical anomalies can be de-weighted before the problem actually occurs. In both AT1 and AT1E the maser frequency drift rates are modeled in the time scale algorithm by using PSFS data from Circular T. AT1E has a frequency drift rate more than a factor of 100 smaller than that of ST15. The BIPM provides time difference data for AT1 relative to International Atomic Time (TAI) \cite{circT}. The BIPM also provides a monthly (typically 30 days, but sometimes 25 or 35 days) average frequency of TAI as measured by primary and secondary frequency standards (PSFS) located in metrology laboratories around the world. We note that PSFS, which is derived from a group of frequency standards, is preferable to TAI, a time scale, as it does not include frequency steps to correct for past time deviations. The frequency difference information is available in Circular T \cite{circT}, which is published once a month. Thus, there is a chain of frequency ratio data from the Yb standard to ST15, to AT1E, to AT1, to TAI, and finally to PSFS. However, the Yb standard is operated only intermittently, and thus there is considerable dead time in the first step of this transfer.

The runs in a month are not distributed evenly. Therefore, the mean time of the runs will not align with the midpoint of the month. Thus, any frequency drift in AT1E will cause a small frequency offset between the mean frequency and what it would be at the midpoint. The frequency drift over the eight month period of interest is less than $7\times10^{-18}$/day. The offsets between mean time and midpoint for each month are approximately randomly distributed about zero, with the largest being -9.1 days and the average being -0.4 days. Thus, the average frequency offset due to AT1E drift is well below $1\times10^{-17}$. Another issue that arose was the presence of an excess number of outliers in the ten (one) second-gated counter (SDR) data. Ninety 5$\sigma$ outliers are found for the entire dataset, where $<1$ are expected from Gaussian statistics. The outliers could potentially bias the mean through an asymmetric distribution. In order to constrain this asymmetry, the Pearson skew is calculated for each individual dataset \cite{Harding2015}. No correlation is found between the skew of a given set and the density of outliers found within that set, leading to a constraint on the role of outliers in data skew at the level of $<3\times10^{-17}$.

Because ST15 operates as a transfer oscillator, its noise does not directly impact the measurement of Yb against PSFS. However, it is necessary to determine the inherent statistical uncertainty arising from the ytterbium clock, frequency comb, and measurement system linked to ST15. This error, denoted here as $\sigma_\mathrm{Yb}$, is generally expected to be very low (e.g., at 1,000 s of averaging, $<1\times10^{-19}$ for the comb \cite{Leopardi2017} and $<1\times10^{-17}$ for the atomic system \cite{Schioppo2016}), but a conservative upper bound is assessed by comparing ST15 noise as measured by the NIST time scale and the Yb/comb system. Measurements of y(Yb$-$ST15) show that the observed Allan deviation, $\sigma_\mathrm{Yb-ST15}=\sqrt{\sigma_\mathrm{Yb}^2+\sigma_\mathrm{ST15}^2}$, is essentially the same as what is observed for ST15 when measured against AT1E, $\sigma_\mathrm{ST15-AT1E}$. Since ST15 contributes to AT1E with weight $w_\mathrm{ST15} \approx$ 10--15\%, it is necessary to account for covariance by using Ref. \cite{Tavella1991}, $\sigma_\mathrm{ST15} = \frac{\sigma_\mathrm{ST15-AT1E}}{(1-w_\mathrm{ST15})^{1/2}}$. More specifically, given the statistical confidence levels for such measurements, the two Allan deviations differ at no more than the $10\%$ level, $\sigma_\mathrm{Yb-ST15} \lesssim 1.1\sigma_\mathrm{ST15}$. This implies that $\sigma_\mathrm{Yb}$ cannot be larger than about one half of the noise of the maser, $\sigma_\mathrm{Yb}=\sqrt{\sigma_\mathrm{Yb-ST15}^2-\sigma_\mathrm{ST15}^2}\lesssim0.5\,\sigma_\mathrm{ST15}$. The minimum measurement interval of AT1 data is 720 s, and the maser noise level at this interval is $2.7\times10^{-15}$. This gives an upper limit to the noise at 720 s of $\sigma_\mathrm{Yb}\approx1.4\times10^{-15}$, or $1.3\times10^{-16}$ at 1 day. A three-corner hat analysis can also be used to get $\sigma_{ST15}$, but it has degraded confidence limits.

The NIST time scale measurement system provides a time difference measurement between the maser ST15 and AT1E on a twelve-minute (720 s) grid. The noise level of a time difference measurement of any length relevant to this study is 1 ps, which contributes a fractional frequency uncertainty of $\sigma_\mathrm{ts}=1.4\times10^{-15}/n$ for a measurement spanning $n$ consecutive twelve-minute bins. The time scale measurement noise, as listed in Table \ref{tab:table1} amounts to $<1\times10^{-17}$ for the full measurement campaign. Another fact to consider is that the y(Yb$-$ST15) measurements do not start and stop on the twelve-minute grid. All data points that fall within a given twelve-minute span are averaged together to project the measurements onto the grid. If the uptime during a given gridpoint is less than 100\%, additional dead time uncertainty, $\sigma_\mathrm{dt}$, is added to this datapoint, increasing the statistical uncertainty associated with this point. Dead time uncertainty is added by using Eq. (2) of Ref. \cite{Yu2007}, which makes use of Eq. (8) of Ref. \cite{Douglas}, and this equation requires knowledge of the noise model of ST15. For an averaging time of 720 s, ST15 can be treated as having a mixed white FM and white PM noise model, and the one-second instabilities of these terms are measured to be $(62\pm4) \times 10^{-15}$ and $(202\pm6) \times 10^{-15}$, respectively. Dead time varies for each twelve-minute gridpoint, depending on uptime. For instance, a unity uptime corresponds to $\sigma_\mathrm{dt}=0$, and an uptime of 10\% corresponds to $\sigma_\mathrm{dt}=8\times10^{-15}$. The total uncertainty of the Yb-maser comparison, as listed in Table \ref{tab:table1}, consists of $\sigma_\mathrm{Yb}$ as well as $\sigma_\mathrm{dt}$ and amounts to $4\times10^{-17}$ for the full measurement.

\begin{figure}
\includegraphics[width=\textwidth,height=\textheight,keepaspectratio]{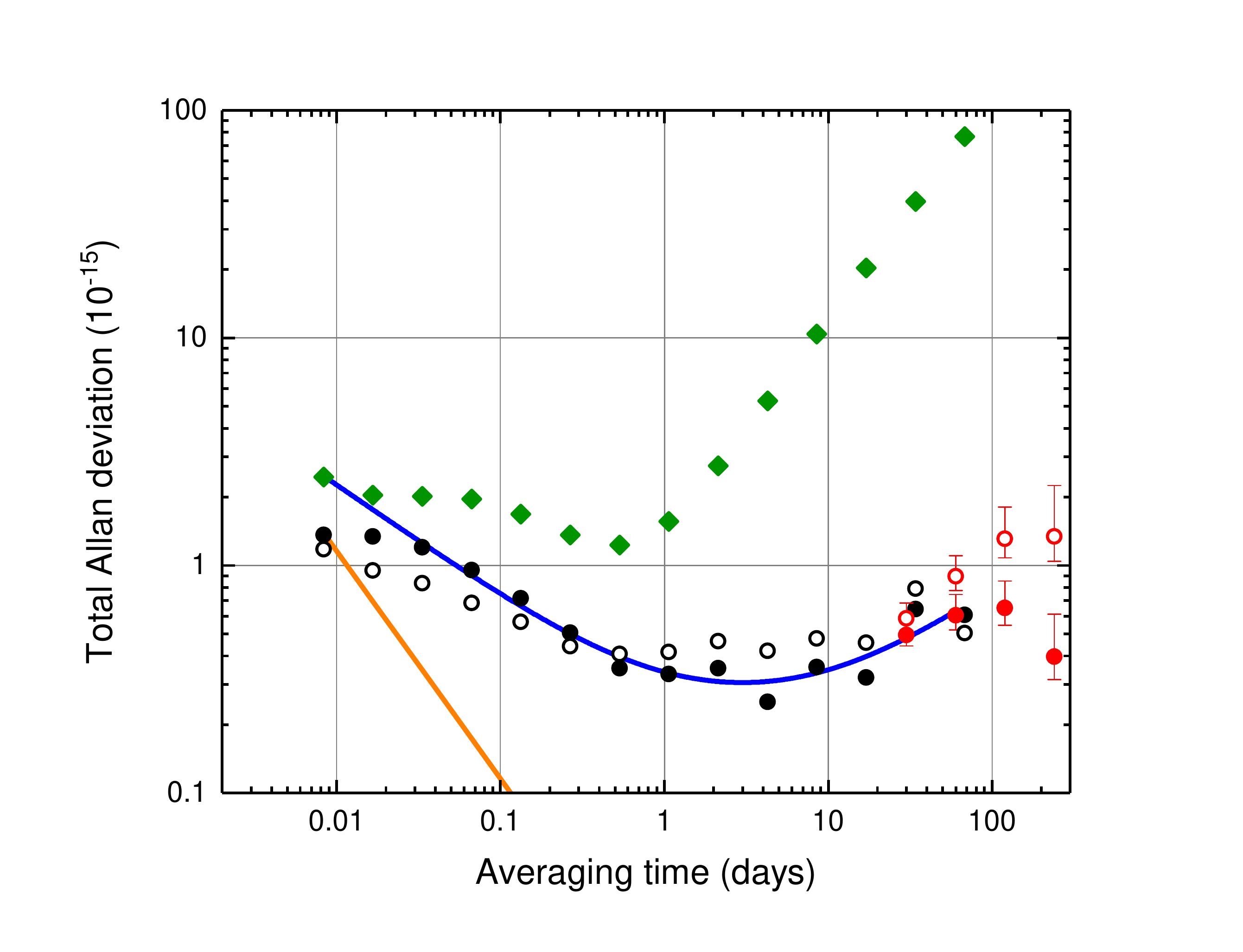}
\caption{\label{tp162} Noise characteristics of relevant frequency references. Filled (empty) circles represent the Allan deviation of AT1E (AT1). Black datapoints are determined by a three-corner hat analysis with hydrogen masers ST11 and ST4, and red datapoints are determined by comparison with Circular T data of PSFS frequencies. Green diamonds are the Allan deviation of ST15, determined by a three-corner hat analysis with ST5 and ST22. AT1E averages down at $>$100 days because y(AT1E$-$PSFS) is used to loosely steer AT1E. The blue line is a fit to a mixed-noise model of AT1E; see main text. The orange line is the time scale measurement uncertainty, $\sigma_\mathrm{ts}$.}
\end{figure}

Thus, we have transformed the intermittent y(Yb$-$ST15) measurements to the twelve-minute-gridded y$_{12\mathrm{\mbox{-}min}}$(Yb$-$ST15), and for each measurement we have a corresponding y$_{12\mathrm{\mbox{-}min}}$(ST15$-$AT1E) measurement with no dead time. From these, we calculate y$_{12\mathrm{\mbox{-}min}}$(Yb$-$AT1E). The next step is to calculate the weighted mean of all the runs that occur in a month to obtain the average frequency difference between the Yb standard and AT1E for a month corresponding to the Modified Julian Dates (MJDs) of the Circular T data.  We will identify this average value as y$_\mathrm{m}$(Yb$-$AT1E). The weighting factor of each contributing 12-minute gridpoint is given by the inverse-square of the uncertainties ($\sigma_\mathrm{Yb}$, $\sigma_\mathrm{ts}$, and $\sigma_\mathrm{dt}$) of that point.

There is considerable dead time uncertainty associated with y$_\mathrm{m}$(Yb$-$AT1E), because the 720-s grid of y$_{12\mathrm{\mbox{-}min}}$(Yb$-$AT1E) is sparsely populated. The dead time uncertainty is calculated using the method of Ref. \cite{Yu2007} for distributed dead time, but the noise characteristics of AT1E must be known in order to calculate this. Fig. \ref{tp162} shows the results of a three-corner hat analysis between AT1E and two independent masers not in AT1E. Confidence limits are not available with the software used to calculate the three-corner hat. The components of the noise model are taken to be white FM, flicker FM, and random walk FM added in quadrature. A fit of the total Allan deviation out to 60 days was performed, and the magnitudes of the total Allan deviation at one day for each component are $(2.25\pm0.11)\times10^{-16}$, $(2.45\pm0.46)\times10^{-16}$, and $(0.750\pm0.085)\times10^{-16}$, respectively. The three-corner hat analysis was performed for a 260 day interval covering approximately the eight month interval of the Yb measurements. This dead time uncertainty represents the largest single source of statistical error for a single month, ranging from $2.0\times10^{-16}$ to $4.6\times10^{-16}$ for the individual months that make up the measurement campaign. Stability plots for AT1 and ST15 are also shown in Fig. 4. The dead time uncertainty is smaller using AT1E than for AT1.

An alternative, but similar, method to that described above was also used. Here the y(Yb$-$AT1E) data was placed in one-day bins, rather than 720 s bins, then combined into monthly averages with dead time uncertainties assigned based on the daily bins. The results for this method are y$_\mathrm{T}$(Yb$-$PSFS) $= (1.7\pm2.1)\times10^{-16}$, with $\chi^2_\mathrm{red}=1.28$, which agrees well with the result reported in the main text, y$_\mathrm{T}$(Yb$-$PSFS) $= (2.1\pm2.1)\times10^{-16}$.

\section{Appendix B: Frequency connection to PSFS}

The frequency differences must now be transferred from AT1E to AT1, as the BIPM only receives timing data from the latter time scale. NIST time scale data are used to obtain y(AT1$-$AT1E) with uncertainties at the $1\times10^{-17}$ level. The fact that there is any uncertainty in y(AT1$-$AT1E) stems from the fact that AT1E and AT1 handle clock anomalies differently \cite{Parker1999}. The next step is to calculate y$_\mathrm{m}$(AT1$-$PSFS) using data from the BIPM. The AT1 time scale is transferred to the BIPM using the hybrid Two-Way Satellite Time and Frequency Transfer/GPS Precise Point Positioning (TWGPPP) frequency transfer protocol \cite{Jiang2009}. The frequency transfer uncertainty in this value can be calculated from Eq. (25) in Ref. \cite{Panfilo2010} and amounts to $3.1\times10^{-16}$, $2.6\times10^{-16}$, and $2.3\times10^{-16}$ for months with lengths of 25, 30, and 35 days, respectively, where we have used the type A uncertainty of 0.4 ns for TAI-AT1 in Circular T \cite{circT}. The transfer uncertainty represents the second largest source of statistical uncertainty, after dead time uncertainty. It is also necessary to transform the clock frequency into the reference surface of the geoid by correcting for the relativistic redshift. This is done with an uncertainty of $6\times10^{-18}$ using the geopotential determination from a recent geodetic survey \cite{Pavlis2017}. Each month the BIPM publishes values for y$_\mathrm{m}$(AT1$-$TAI) as well as y$_\mathrm{m}$(TAI$-$PSFS) \cite{circT}. Thus, we have finally,

\begin{widetext}
y$_\mathrm{m}$(Yb$-$PSFS)=y$_\mathrm{m}$(Yb$-$AT1E)+y$_\mathrm{m}$(AT1E$-$AT1)+y$_\mathrm{m}$(AT1$-$TAI)+y$_\mathrm{m}$(TAI$-$PSFS),
\end{widetext}
where all y$_\mathrm{m}$ are for the same time interval (25, 30, or 35 days). The uncertainty budget is listed in Table \ref{tab:table1}, displaying the uncertainties associated with the month of March, as well as for the full campaign.

The type A (statistical) and type B (systematic) uncertainties for each primary and secondary frequency standard reporting are given in Circular T for a specific time interval. In addition to the individual standards, a total uncertainty for the weighted mean of y(TAI$-$PSFS) is also given. This total uncertainty is calculated from the type A and B uncertainties of the individual standards, as well as the uncertainty in transferring frequency information from each standard to TAI. During the course of the eight month measurement, ten separate standards contributed to PSFS: two Cs thermal beam clocks (PTB-CS1 and PTB-CS2), seven Cs fountains (PTB-CsF1, PTB-CsF2, SYRTE FO1, SYRTE FO2, SU-CsF2, IT-CsF2, and NIM5) \cite{Guena2012,Domnin2013,Levi2014,Fang2015,Weyers2018}, and a single Rb fountain (SYRTE FORb) \cite{Guena2014}, with between five and eight of these being present in any given month. The SYRTE and PTB Cs fountains were present most often and have the most weight.

An alternative approach to that described above was also used to compare Yb against PSFS via the chain of Yb$\rightarrow$AT1$\rightarrow$TAI$\rightarrow$PSFS. Instead of processing the data monthly, this approach treats each optical clock run individually and then processes the full-campaign data all together. First, we determine the frequency difference y$_\mathrm{r}$(Yb$-$AT1) for a given run. Then, this value is added to y$_\mathrm{avg}$(AT1$-$TAI), the average frequency difference between the NIST time scale and TAI, which is published on a 5-day grid. This step adds long-term noise from the frequency transfer techniques from NIST to the BIPM. Each of the runs, y$_\mathrm{r}$(Yb$-$TAI), are averaged together to yield y$_\mathrm{T}$(Yb$-$TAI). However, since dead-time uncertainty has not been yet accounted for, the points are overscattered, and the uncertainty of the mean is expanded by $\sqrt{\chi^2_\mathrm{red}}=1.6$. The resultant mean is added to y$_\mathrm{T}$(TAI$-$PSFS), published by BIPM. This approach yields the frequency difference y$_\mathrm{T}$(Yb$-$PSFS) $=(2.0\pm2.0)\times10^{-16}$, in good agreement with the results in the main text, y$_\mathrm{T}$(Yb$-$PSFS) $=(2.1\pm2.1)\times10^{-16}$. This good agreement using a very different analysis method demonstrates that the results in the main text are robust against alternative methods of analysis.

\section{Appendix C: Yb/Rb ratio and further loop closures}

\begin{figure}
\includegraphics[width=\textwidth,height=\textheight,keepaspectratio]{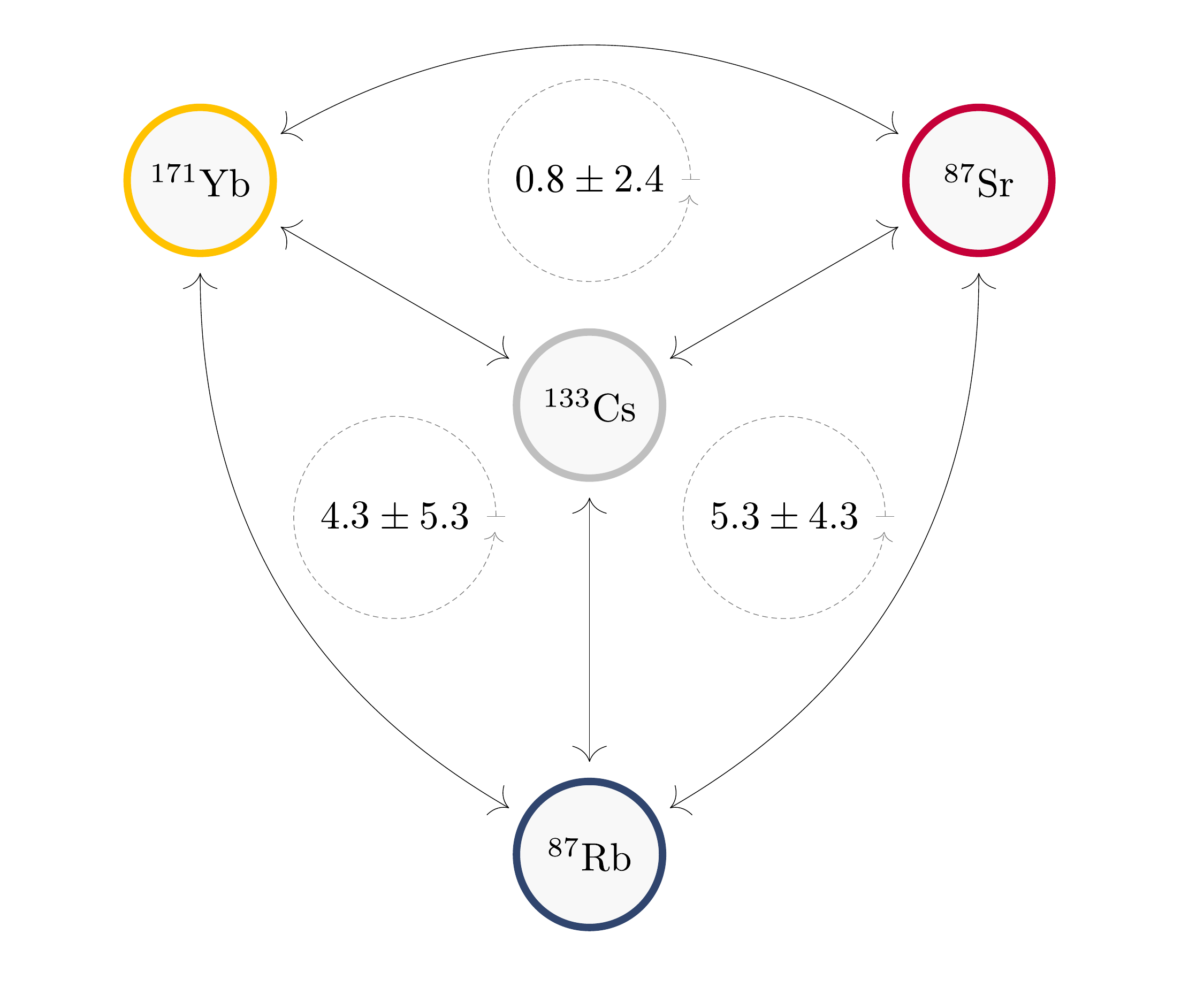}
\caption{\label{rbloops} Diagram representing three loop closures involving optical frequency standards and microwave frequency standards. The numbers represent the loop misclosure (as parts in $10^{16}$), as defined in the main text. Comparisons are represented by connections between the standards. The contributing Yb absolute frequency measurements were from NIST, INRIM, KRISS, NMIJ; Sr absolute measurements from PTB, SYRTE, NICT, JILA, NMIJ, NIM, UT; Rb absolute measurements from NPL, SYRTE; Yb/Sr optical ratios from RIKEN, PTB/INRIM, NMIJ, KRISS/NICT; Yb/Rb ratio from NIST; Sr/Rb ratio from SYRTE.}
\end{figure}

A single secondary representation of the second, the $^{87}$Rb hyperfine transition, contributed to PSFS over this eight-month period \cite{Guena2014}, though the relatively high uncertainty ($6\times10^{-16}$) given by the BIPM to this transition meant that its impact was negligible relative to the Cs fountains \cite{Riehle2018}. However, since the frequency difference from TAI is given for each individual standard, it is possible to determine y$_\mathrm{m}$(Yb$-$SYRTE FORb), allowing a determination of the Yb-Rb frequency ratio. The SYRTE FORb frequency standard contributed for three of the eight months, for a combined total Yb-Rb runtime of 3.3 days. The type A uncertainty for the three months is $3.3\times10^{-16}$, and the type B uncertainty of the standard was $2.7\times10^{-16}$, leading to a total uncertainty of $4.3\times10^{-16}$. The ratio is found to be $\nu_\mathrm{Yb}/\nu_\mathrm{Rb}=$75 833.197 545 114 192(33). This agrees well with $\nu_\mathrm{Yb}^\mathrm{BIPM17}/\nu_\mathrm{Rb}^\mathrm{BIPM17}=$75 833.197 545 114 196(59). For the Yb/Rb measurements, $\chi_\mathrm{red}^2=0.72$.

 \begin{table}
 \caption{\label{tab:table3} Loop closures performed on the Cs-X-Y-Cs loop, where X and Y are frequency standards.}
 \begin{ruledtabular}
 \begin{tabular}{llcccc}
X & Y & X/Cs ref. & Y/Cs ref. & X/Y ref. & Misclosure ($10^{-16}$)\\
\colrule
$^{171}$Yb & $^{87}$Sr & this work, \cite{Lemke2009,Kohno2009,Yasuda2012,Park2013,Kim2017,Pizzocaro2017} & \cite{Ludlow2006,LeTargat2006,Boyd2007,Baillard2008,Gretchen2008,Hong2009,Falke2011,Yamaguchi2012,LeTargat2013,Akamatsu2014a,Falke2014,Tanabe2015,Lin2015,Lodewyck2016,Grebing2016,Hachisu2017,Hachisu2017a},  & \cite{Akamatsu2014,Takamoto2015,Nemitz2016,Grotti2018,Akamatsu2018,Fujieda2018} & $0.8\pm2.4$\\
$^{171}$Yb & $^{87}$Rb & " & \cite{circT,Ovchinnikov2015,Guena2017} & this work & $4.3\pm5.3$\\
$^{87}$Sr & $^{87}$Rb &\cite{Ludlow2006,LeTargat2006,Boyd2007,Baillard2008,Gretchen2008,Hong2009,Falke2011,Yamaguchi2012,LeTargat2013,Akamatsu2014a,Falke2014,Tanabe2015,Lin2015,Lodewyck2016,Grebing2016,Hachisu2017,Hachisu2017a},   & " & \cite{Lodewyck2016} & $5.3\pm4.3$\\
$^{87}$Sr & $^{199}$Hg & " & \cite{McFerran2012,Tyumenev2016} & \cite{Yamanaka2015,Tyumenev2016} & $-1.6\pm4.5$\\
$^{87}$Sr & $^{88}$Sr & " & \cite{Baillard2007, Morzynski2015} & \cite{Katori2008,Takano2017,Origlia2018} & $-21\pm40$\\
$^{87}$Sr & $^{40}$Ca$^+$ & " & \cite{Chwalla2009,Matsubara2012,Huang2012,Huang2016} & \cite{Matsubara2012} & $48\pm50$\\
$^{199}$Hg & $^{87}$Rb & \cite{McFerran2012,Tyumenev2016} & \cite{circT,Ovchinnikov2015,Guena2017} & \cite{Tyumenev2016} & $4.8\pm6.1$\\
$^{171}$Yb$^+$E2 & $^{171}$Yb$^+$E3 & \cite{Tamm2009,Webster2010,Tamm2014,Godun2014} & \cite{Hosaka2009,Huntemann2012,King2012,Huntemann2014,Godun2014,Godun2018} & \cite{Godun2014} & $-14.0\pm5.6$\\
$^{27}$Al$^+$ & $^{199}$Hg$^+$ & \cite{Rosenband2007} & \cite{Stalnaker2007} & \cite{Rosenband2008} & $-57\pm54$\\
 $^{40}$Ca & $^{199}$Hg$^+$ & \cite{Degenhardt2005,Wilpers2007} & " &  \cite{Vogel2001} & $-100\pm760$\\
\end{tabular}
 \end{ruledtabular}
 \end{table}

The measurement of the Yb/Rb ratio allows the determination of two additional frequency loop misclosures; see Fig. \ref{rbloops}. The Rb absolute frequency is determined as $\nu_\mathrm{Rb}=$ 6 834 682 610.904 311 5(16) Hz, by a weighted average of three previous measurements \cite{circT,Ovchinnikov2015,Guena2017}. The displayed Yb/Rb ratio is from this work, and the Sr/Rb ratio is from \cite{Lodewyck2016}. In each of the three loops examined (Yb/Sr, Yb/Rb, Sr/Rb), misclosures are consistent with zero at an uncertainty smaller than $6\times10^{-16}$. In Table \ref{tab:table3}, we compile a list of all loops that have been closed with Cs as one of the three points. As expected from statistics, three of the ten loops are at least $1\sigma$ from zero. Only a single loop has a misclosure $>2\sigma$ from zero.

\section{Appendix D: Constraints on variations of constants from Yb and Sr absolute frequency measurements}

\begin{figure}
\includegraphics[width=\textwidth,height=\textheight,keepaspectratio]{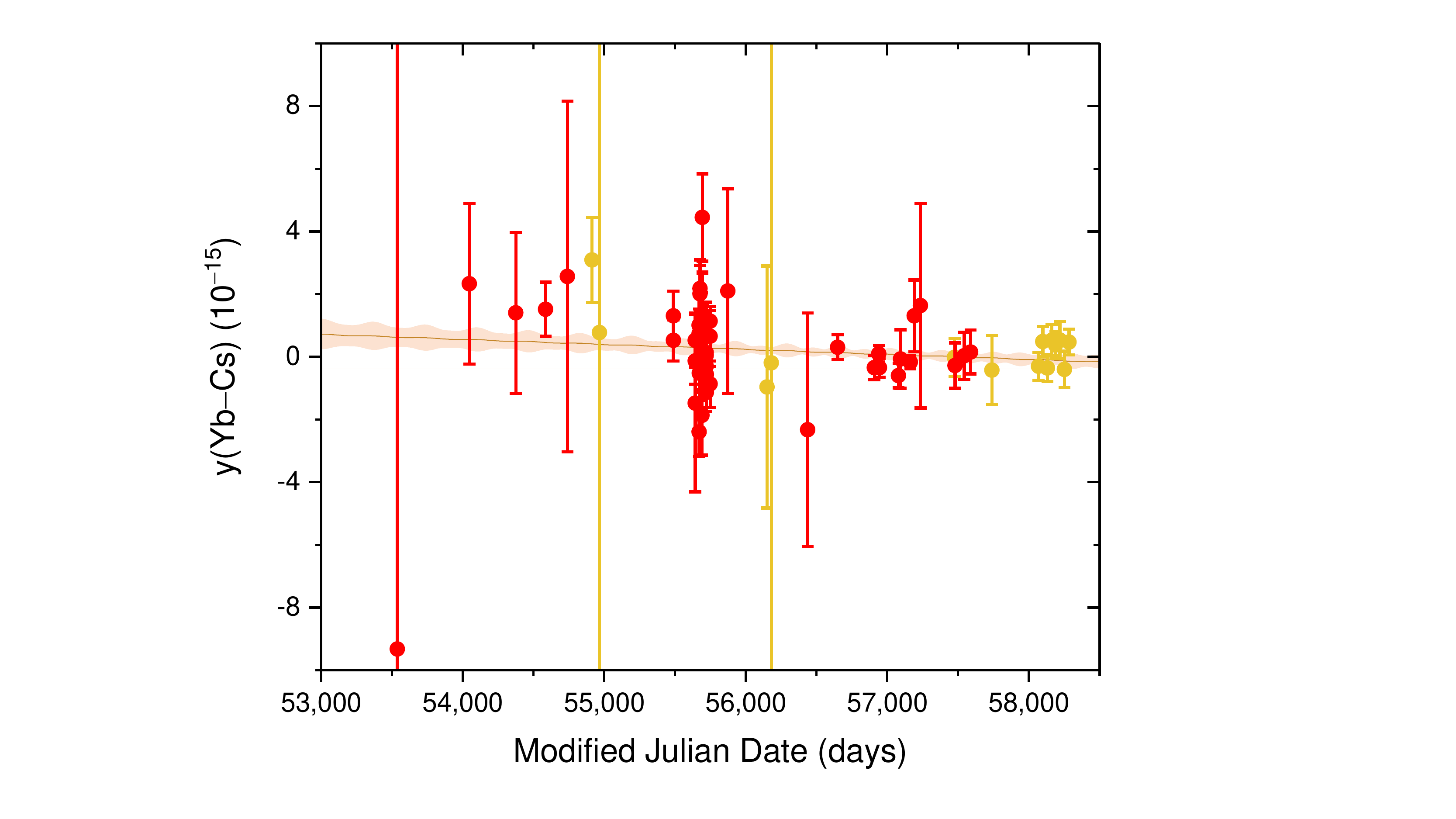}
\caption{ \label{ybplussr} A summary of this work, as well as six other Yb absolute frequency measurements (yellow) \cite{Lemke2009,Kohno2009,Yasuda2012,Park2013,Kim2017,Pizzocaro2017}, and seventeen Sr measurements (red) \cite{Ludlow2006,LeTargat2006,Boyd2007,Baillard2008,Gretchen2008,Hong2009,Falke2011,Yamaguchi2012,LeTargat2013,Akamatsu2014a,Falke2014,Tanabe2015,Lin2015,Lodewyck2016,Grebing2016,Hachisu2017,Hachisu2017a}. Sr measurements are multiplied by the Yb/Sr ratio and the error bars are expanded as descibed in the main text. The fit function (orange) constrains possible variation of the Yb/Cs ratio, and the shaded region represents the 1$\sigma$ uncertainty of the fit.}
\end{figure}

To tighten the contraints on the possible variations of fundamental constants, we employ a multi-species analysis that can be used to link absolute frequency measurements of different types of atomic clocks with optical ratios that have been measured with sufficient precision. We link our results with prior absolute measurements of the $^{171}$Yb and $^{87}$Sr transition frequencies. First, we multiply the Sr transition frequency by the well-known ratio, $\mathcal{R}_\mathrm{opt}^\mathrm{avg}=$ 1.207 507 039 343 337 768(60), derived from Refs \cite{Akamatsu2014,Takamoto2015,Nemitz2016,Grotti2018,Akamatsu2018,Fujieda2018} as described in the main text. The ratio is known with a fractional uncertainty of $\sigma_\mathrm{stat}=5.6\times10^{-17}$. However, care must be taken during this step due to the non-zero sensitivity of the Yb/Sr ratio to variation of the fine structure constant (N.B.: $\Delta K_\mathrm{Yb,Sr}^{X_q} =\Delta K_\mathrm{Yb,Sr}^{\mu}=0)$. Prior analyses have constrained $\frac{\dot\alpha}{\alpha}$ to be consistent with zero at the level of $2\times10^{-17}$/yr \cite{Godun2014,Huntemann2014}, and $k_\alpha(\frac{\Delta\Phi}{\mathrm{c}^2})$ to be consistent with zero at the level of $1.7\times10^{-17}$ \cite{Dzuba2017}. Multiplying these numbers by $\Delta K_\mathrm{Yb,Sr}^\alpha=0.25$, we obtain a constraint on the time-variation of the Yb-Sr ratio of $<5\times10^{-18}$/yr and a constraint on gravitational potential-variation of $\sigma_\mathrm{grav}=4\times10^{-18}$. To account for a possible drift in $\alpha$ in a conservative fashion, we multiply the former number by twelve years, the entire duration of the record of absolute frequency measurements, leading to $\sigma_\mathrm{time}=6\times10^{-17}$. Finally, we expand the error bars of each of the rescaled Sr absolute frequency measurements by $\sqrt{\sigma_\mathrm{stat}^2+\sigma_\mathrm{grav}^2+\sigma_\mathrm{time}^2}=8.2\times10^{-17}$. For all Sr measurements, this increased uncertainty had only a slight impact upon the initial error bar, the lowest of which was $2.6\times10^{-16}$, or $2.7\times10^{-16}$ after rescaling \cite{Grebing2016}.

The seven Yb measurements \cite{Lemke2009,Kohno2009,Yasuda2012,Park2013,Kim2017,Pizzocaro2017} and 17 rescaled Sr measurements \cite{Ludlow2006,LeTargat2006,Boyd2007,Baillard2008,Gretchen2008,Hong2009,Falke2011,Yamaguchi2012,LeTargat2013,Akamatsu2014a,Falke2014,Tanabe2015,Lin2015,Lodewyck2016,Grebing2016,Hachisu2017,Hachisu2017a}  are displayed in Fig. \ref{ybplussr}. For Refs. \cite{LeTargat2013,Lodewyck2016,Hachisu2017a}, measurements took place over many months, and so individual runs were extracted from the published work, rather than simply using the average. In order to constrain variation of constants arising from both time and gravitational potential, we fit the function y(Yb$-$Cs) $= A\mathrm{ cos}(2\pi\frac{t-t_0}{1\,\mathrm{yr}})+\dot{y}t+y_0$, where t is the date. $A$, $\dot y$, and $y_0$ are free parameters. We determine $A=(0.1\pm1.5)\times10^{-16}$ and $\dot y=(-4.9\pm3.6)\times10^{-17}/$yr. The Yb/Cs measurements and rescaled Sr/Cs measurements are sensitive to variations of $\alpha$, $X_q$, and $\mu$.

Using the values in Table \ref{tab:table2} and the amplitude of the sinusoidal fit, we find $k_{\mu}=(-0.19\pm0.94)\times10^{-6}$, indicating a constraint a factor of 1.5 below the constraint derived from our measurements alone. We also use the linear fit to constrain the drift of $\mathrm{\mu}$ by using the equation, $\dot y=\sum\limits_\epsilon \Delta K_{\mathrm{Yb,Cs}}^\epsilon \frac{\dot\epsilon}{\epsilon}$. We note that, using the values of $\dot \alpha$ and $\dot X_q$ from Ref. \cite{Godun2014} and Ref. \cite{Guena2012}, $\Delta K_{\mathrm{Yb,Cs}}^\alpha (\frac{\dot\alpha}{\alpha})=(1.8\pm5.3)\times10^{-17}/\mathrm{yr}$ and $\Delta K_{\mathrm{Yb,Cs}}^{\mathrm{X_q}} (\frac{\dot X_q}{X_q})=(-1.4\pm0.9)\times10^{-17}$/yr. Solving for $\mu$, we find that $\frac{\dot\mu}{\mu}=(5.3\pm6.5)\times10^{-17}$/yr. This represents an improvement of almost a factor of two below the previous constraint, $\frac{\dot\mu}{\mu}=(-2\pm11)\times10^{-17}$/yr \cite{Godun2014}.




\begin{acknowledgments}
This work was supported by NIST, NASA Fundamental Physics, and DARPA QuASAR. The authors thank J. C. Bergquist and N. Ashby for their careful reading of the manuscript. Initial development of the AT1 time scale was facilitated by J. Levine. This work is a contribution of the National Institute of Standards and Technology/Physical Measurement Laboratory, an agency of the U.S. government, and is not subject to U.S. copyright.
\end{acknowledgments}

\begin{flushleft}
$^*$Permanent address: State Key Laboratory of Advanced Optical Communication Systems and Networks, Institute of Quantum Electronics, School of Electronics Engineering and Computer Science, Peking University, Beijing 100871, China

$^\dagger$Permanent address: Niels Bohr Institute, University of Copenhagen, Blegdamsvej 17, 2100 Copenhagen, Denmark

$^\ddagger$Present address: Georgia Tech Research Institute, Atlanta, GA 30332, USA

$^\mathsection$tom.parker@nist.gov

$^\mathparagraph$tara.fortier@nist.gov

$^\|$andrew.ludlow@nist.gov
\end{flushleft}

\bibliography{paper}

\end{document}